# The influence of repressive legislation on the structure of a social media network

Marianne Marcoux[(a)] and David Lusseau

*Institute of Biological and Environmental Sciences, University of Aberdeen - Zoology Building, Tillydrone Avenue, AB24 2TZ, Aberdeen, UK*

**Abstract** – Social media have been widely used to organize citizen movements. In 2012, 75% university and college students in Quebec, Canada, participated in mass protests against an increase in tuition fees, mainly organized using social media. To reduce public disruption, the government introduced special legislation designed to impede protest organization. Here, we show that the legislation changed the behaviour of social media users but not the overall structure of their social network on Twitter. Thus, users were still able to spread information to efficiently organize demonstrations using their social network. This natural experiment shows the power of social media in political mobilization, as well as behavioural flexibility in information flow over a large number of individuals.

**Introduction.** – Social media are powerful tools that can be used to spread information and organize protest [1-4]. The potential of social media to facilitate citizen movements is widely known [5] and their use is banned in some authoritarian states [6,7]. For example, social media played an important role in organizing demonstrations against the government in Egypt in 2011 [1]. During Moldova's Revolution in April 2009, new internet-based information and communication technologies allowed the coordination of the revolution [2].

Social media also provides extensive data sets for social science research [8]. Social media data sets create new opportunities to investigate human behaviour such as the spread of information and opinions, as well as the formation of groups. For example, a study using the Facebook social network showed that the political opinions of individuals were strongly influenced by the opinions of their Facebook friends, and the friends of their Facebook friends [4]. The large size of the Facebook network was critical for allowing Ugander *et al.* to unravel subtle mechanisms that underlies social contagion in this example [9]. In addition, policy makers have identified the importance of regulating online activity (e.g. access to data, usage of social media to organize social demonstration) [5]. However, it is becoming apparent to policy makers that there is a paucity of data available to them to guide decision making and measure the impact their regulation [5]. Thus, case-studies involving large data sets are required not only to develop analysis techniques, but also to gain insight into the response of people to policy decision.

In the winter of 2012, 75% (about 300 000) students of colleges and universities in the province of Quebec, Canada, went on strike to protest against an announced increase in tuition fees. Social media such as Twitter were used to organize demonstrations [10]. On May 18, after 14 weeks of the strike, the Quebec government passed special emergency legislation (Bill-78) [11] with the objective of reducing protests. In short, under Bill 78, any organizer of a demonstration involving more than 50 persons had to provide the local police with an itinerary of the demonstration not later than 8 hours before the event and ensure that the demonstration followed the itinerary. Severe fines were planned for individuals or organizations who did not follow Bill-78. There were also discussions in the Parliament that individuals using social media to organize or transfer information about illegal demonstrations could also be fined [12].

Here, we provide a case study where legislation aiming at stopping a protest movement might affect the behaviour of social media users and the structure of their interaction network. We compare the structure of the interaction network before and after the passage of Bill 78 as well as network measures of individual Twitter users.

**Material and methods.** – Given the wide-usage of the social media Twitter during the student strike to organize demonstrations, the aim of Bill-78 to inhibit student protest, and discussions in Parliament on the applicability of Bill-78 to Twitter users, we investigated the impact of Bill-78 on the behaviour of Twitter users and on the structure of the social network emerging from their interactions. Here, the social network describes a set of nodes, people using Twitter, and their patterns of interactions, their exchanges of "tweets". The network was directed because it reflected the direction of the tweets (from the sender to the addressee) and weighted to represent the rate exchanges of tweets between two users (tweets/day). We built a network for before (Feb 12-May 17, 105 005 tweets) and after the passage of Bill-78 (May 18-June 4, 89 589 tweets).

A search of Twitter (http://twitter.com) was performed to find tweets including the following hashtags (keywords) associated with the student strike from Feb 12 to June 4 (the date of the beginning of our analysis); #ggi , #manifencours, #casseroles and #non1625. For each tweet, the publication date, sender user name, hashtags, and addressee of the tweet were noted. To be included in our Twitter networks, a user needed to have written at least one tweet addressed to another user who also wrote tweets about the strike. We identified most influential

(a) marianne.marcoux@gmail.com

students associations and leaders within Twitter users with the top 100 eigenvector centrality values.

**Results and discussion.**-Firstly, we investigated the effect of Bill-78 on the activity rate on Twitter. While the number of tweets posted per day was higher after Bill-78, the rate of increase in tweets posted per day dropped after the passage of Bill-78 (May 18, figure 1). The best fitted piecewise regression model included break points at April 17 and May 17 (Anova, $F_{4, 106}$= 45.96, p<0.001, compared to model with no breaks, break points were estimated by looking at the residual standard error of models with all possible break points, figure 1). Bill-78 affected the rate of daily activity on Twitter since there was a positive change in the number of tweets/day before Bill-78 (12 extra tweets/day/day before April 17 and 29 extra tweets/day/day after April 17) while the rate of tweets/day decreased after the passage of Bill-78 (241 less tweets/day/day).

Secondly, we investigated the change in behaviours of Twitter users after the passage of Bill-78 using a paired design. Eigenvector centrality measures the connectivity of a user as well as the connectivity of its network neighbours; a user can acquire a high connectivity from the number and weight of his interactions or from being connected to highly connected users [13]. Eigenvector centrality can be easily calculated for large weighted network [14]. The average eigenvector centrality values of the 2836 Twitter users who wrote about the strike both before and after the passage of Bill-78 Twitter users decreased after Bill-78 (paired permutation test with 9999 permutations, p=0.001). This decrease was significantly higher for the most influential student associations than for the rest of the Twitter users (permutation test with 9999 permutations, p< 0.001, figure 2). Thus, Bill-78 decreased the average connectivity of network users resulting in a decentralization of information transfer. After Bill-78, the control of information was better distributed among Twitter user. Decentralization has also been documented as the effect of intensive law-enforcement on a criminal network [15]. The clustering coefficient measures the connectivity of the neighbours of a user to each other, and provide an indication of cliquishness [16]. Because our networks were directed, we obtain two measures for the clustering coefficient, the in-clustering coefficient for the received tweets and the out-clustering coefficient for the tweets sent. While the out-clustering coefficient increased (p=0.004), the in-clustering coefficient did not significantly change (p=0.699). This increase in out-clustering coefficient was larger for the student associations than for the rest of the Twitter users (p= 0.037; figure 2) and indicates an increase in the cliquishness of the network of tweet sent. As a consequence, information was more likely to stay within the cliques of users. This may have resulted after Bill-78 because students were more cautious of the recipients of their information about protests.

Lastly, we investigated if these changes in behaviours affected the overall structure of the Twitter network (figure 3). Both networks before and after the passage of Bill-78 followed a scale-free distribution [17,18] for both the in-strength (or weighted in-degree, another measure of connectivity calculated by the sum of the rates of the received tweets of an individual) and the out-strength (or weighted out-degree, sum of the rates of the tweets sent; figure 4). Thus, a few Twitter users had very high rates of tweet exchanges while most of the users had low rates of exchanges. Before and after Bill-78, the networks of Twitter users were hierarchical because the clustering coefficients of individuals were distributed following a power law in relation to their strengths [19]. In order words, individuals with a low connectivity tended to be found in highly interconnected cliques while individuals with high connectivity were positioned between cliques. There was no significant changes in the hierarchy of the network before and after the passage of Bill-78 (pre-Bill models: $C_{in}(k_{in}) \sim k_{in}^{-0.48 \pm 0.04 \text{ s.e.m.}}$, $C_{out}(k_{out}) \sim k_{out}^{-0.29 \pm 0.03 \text{ s.e.m.}}$; post-Bill models: $C_{in}(k_{in}) \sim k_{in}^{-0.47 \pm 0.05 \text{ s.e.m.}}$, $C_{out}(k_{out}) \sim k_{out}^{-0.36 \pm 0.03 \text{ s.e.m.}}$). In hierarchical networks, small groups of highly connected individuals are organized in a hierarchical manner into increasingly large groups, while maintaining a scale-free topology [19].

**Conclusion.**-The passage of Bill-78 changed some of the interactions among Twitter users but not the structure of their social network emerging from these interactions. While Twitter users were more likely to restrict their tweet exchanges to within their cliques after Bill-78, the scale-free and hierarchical structure of the network after the passage of Bill-78 still allowed for information flow about demonstration [20]. Thus, individuals were still able to organize mass social demonstrations [21] thanks to the overall structure of their interaction network [22]. In conclusion, our results show that social network structures can be resilient to changes in the behaviour of individuals that compose those networks and are effective at maintaining information flow under a wide range of interaction behaviours [2,22].

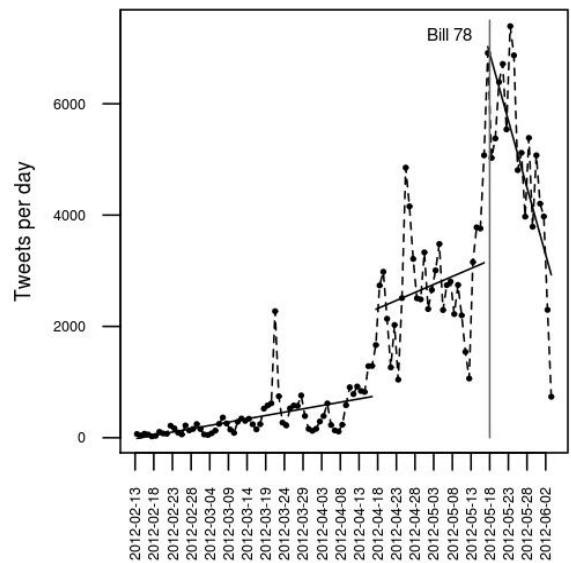

**Fig. 1:** Rate of tweets published per day related to the 2012 Quebec student strike. The vertical line shows the date of the passage of Bill-78 and the solid lines represent the piecewise regression ($R^2$ = 0.88, p < 0.001).

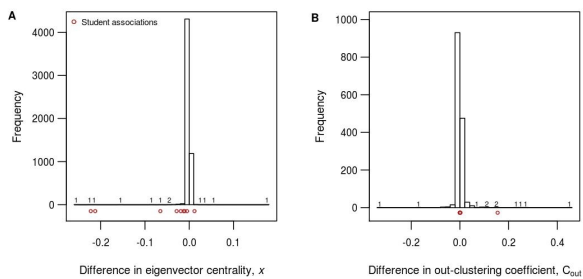
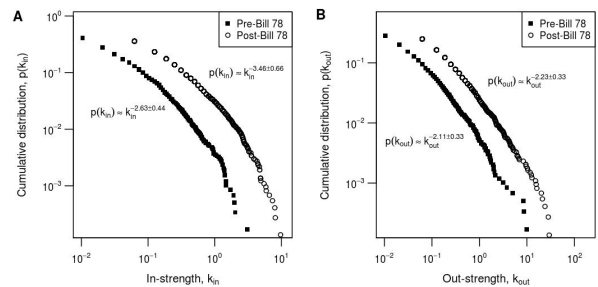

**Fig. 2:** (Colour on-line) Histogram of paired differences in A) eigenvector centrality values, and B) out-clustering coefficient before and after the passage of Bill-78. Values for the 9 most influential student association are represented by the red circles.

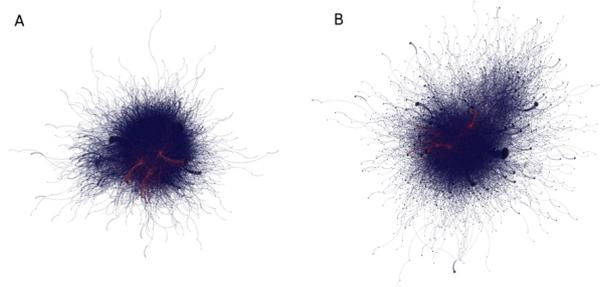

**Fig. 3:** (Colour on-line) Giant component (largest connected cluster) of the network of tweet exchanges related to the student strike A) before (n= 5895) and B) after (n=7342) the passage of Bill-78. Node sizes are relative to in-strength. The 9 student associations nodes and connections are in red.

**Fig. 4:** Distribution of A) in- and B) out-strength of the network of social media users before and after Bill-78 showing scale-free network structure. Models are calculated from [18] and ± represent standard deviation.

***

We want to thank O. H. Beauchesne for making the Twitter data publicly available. Q. E. Fletcher and three anonymous reviewers gave constructive comments on the manuscript. D. L. acknowledge the support of the MASTS pooling initiative (The Marine Alliance for Science and Technology for Scotland) in the completion of this study. MASTS is funded by the Scottish Funding Council (grant reference HR09011) and contributing institutions. M. M. funding was provided by the Fonds de Recherche Nature et Technologie Québec. Thanks to the University of Aberdeen for the usage of the RINH/BioSS Beowulf cluster.

(a) marianne.marcoux@gmail.com